\documentclass[aps,prl,twocolumn,superscriptaddress]{revtex4-1}

\usepackage{amsmath}
\usepackage{amssymb}
\usepackage{braket}
\usepackage{color}
\usepackage{graphicx}
\usepackage{epsfig}
\usepackage[outdir=./Figures/]{epstopdf}
\usepackage[implicit=false,hidelinks]{hyperref}
\usepackage[normalem]{ulem}

\graphicspath{{Figures/}}

\DeclareMathOperator{\Tr}{Tr}

\begin{document}
	
	\title{Superradiant Switching, Quantum Hysteresis, and Oscillations \\ in a Generalized Dicke Model}
	
	\author{Kevin C. Stitely}
	\email{kevin.stitely@auckland.ac.nz}
	\affiliation{Dodd-Walls Centre for Photonic and Quantum Technologies, New Zealand}
	\affiliation{Department of Mathematics, University of Auckland, Auckland 1010, New Zealand}
	\affiliation{Department of Physics, University of Auckland, Auckland 1010, New Zealand}
	
	\author{Stuart J. Masson}
	\affiliation{Department of Physics, Columbia University, New York, New York 10027, USA}
	\author{Andrus Giraldo}
	\affiliation{Dodd-Walls Centre for Photonic and Quantum Technologies, New Zealand}
	\affiliation{Department of Mathematics, University of Auckland, Auckland 1010, New Zealand}
	
	\author{Bernd Krauskopf}
	\affiliation{Dodd-Walls Centre for Photonic and Quantum Technologies, New Zealand}
	\affiliation{Department of Mathematics, University of Auckland, Auckland 1010, New Zealand}
	
	\author{Scott Parkins}
	\affiliation{Dodd-Walls Centre for Photonic and Quantum Technologies, New Zealand}
	\affiliation{Department of Physics, University of Auckland, Auckland 1010, New Zealand}
	
	\date{\today}
	
	\begin{abstract}
		We demonstrate quantum signatures of deterministic nonlinear dynamics in the transition to superradiance of a generalized open Dicke model with different coupling strengths for the co- and counter-rotating light-matter interaction terms. A first-order phase transition to coexisting normal and superradiant phases is observed, corresponding with the emergence of switching dynamics between these two phases, driven by quantum fluctuations. We show that this phase coexistence gives rise to a hysteresis loop also for the quantum mechanical system. Additionally, a transition to a superradiant oscillatory phase can be observed clearly in quantum simulations. 
	\end{abstract}
	
	\maketitle
	
	\textit{Introduction.}---Atoms in a dilute gas radiate light independently of one another through spontaneous emission, whereby the number of excitations decreases exponentially \cite{loudon_quantum_1983,scully_quantum_1997}. In contrast, a large, tightly-confined ensemble of atoms has its emission enhanced by coherence stored in the atoms and emits strong pulses of radiation over much shorter timescales than independent atoms \cite{gross_superradiance:_1982}. This phenomenon, known as \emph{superradiance}, was first suggested in a seminal paper by Dicke in 1954 \cite{dicke_coherence_1954}. Hepp and Lieb later showed that an ensemble coupled to a single quantized radiation mode will undergo a quantum phase transition to superradiance in the steady state in a model now known as the Dicke model \cite{hepp_equilibrium_1973,hepp_superradiant_1973,wang_phase_1973}.
	
	This phase transition is often studied in the semiclassical (mean-field) regime where, if the number of atoms is sufficiently high, the role of quantum fluctuations can be neglected. In this case, the dynamics of the system are governed by a system of nonlinear differential equations, and the transition to superradiance emerges via a pitchfork bifurcation \cite{kirton_introduction_2019} which breaks the system's $\mathbb{Z}_{2}$ symmetry.
	
	Recently, the Dicke model has had experimental realizations in both Bose--Einstein-condensate and trapped-ion systems \cite{baumann_dicke_2010,baumann_exploring_2011,hamner_dicke-type_2014,safavi-naini_verification_2018}. Another realization was proposed by Dimer \emph{et al.} \cite{dimer_proposed_2007} using an ensemble of four-level atoms confined to an optical cavity with atom-light interactions generated by stimulated Raman transitions between atomic ground states. This proposal allows for two different coupling strengths for the co- and counter-rotating terms of the Hamiltonian. This possibility, which we refer to as unbalanced coupling, was recently realized experimentally \cite{zhiqiang_nonequilibrium_2017,zhang_dicke_2018} and was shown to lead to additional phases, including an oscillating superradiant phase.
	
	Motivated by these experiments, a recent analysis of the semiclassical model found a diverse set of complex nonlinear behavior \cite{stitely_nonlinear_2020}, including two types of chaotic dynamics in the superradiant domain. In this Letter, we explore the manifestation of deterministic nonlinear dynamics in the fully quantum mechanical system. We show that the appearance of a first-order quantum phase transition leads to a state in which both the normal and superradiant phases coexist, which has also been shown in a similar model with atomic dissipation \cite{gelhausen_dissipative_2018}. We then simulate the quantum dynamics with a stochastic Schr\"{o}dinger equation to demonstrate switching between these phases triggered by quantum fluctuations. We also identify the superradiant oscillatory phase, emerging due to a Hopf bifurcation in the semiclassical model, in the fully quantum mechanical model. In this way, our results yield insight into quantum phase transitions and the quantum--classical correspondence.

	
	\textit{Model.}---The unbalanced Dicke model consists of an ensemble of $N$ two-level atoms confined to an optical cavity with a single field mode, with coupling strengths $\lambda_{-}$ for the co-rotating terms and $\lambda_{+}$ for the counter-rotating terms. The Hamiltonian is ($\hbar=1$)
	\begin{align}\label{eqn:hamiltonian}
	\hat{H} =&\  \omega \hat{a}^{\dagger}\hat{a} + \omega_0 \hat{J}_{z} + \frac{\lambda_{-}}{\sqrt{N}}\left( \hat{a}\hat{J}_{+} + \hat{a}^{\dagger}\hat{J}_{-} \right) \nonumber \\&+ \frac{\lambda_{+}}{\sqrt{N}}\left( \hat{a}\hat{J}_{-} + \hat{a}^{\dagger}\hat{J}_{+} \right),
	\end{align}
	where $\hat{a}$ is the cavity field annihilation operator, $\omega$ is the effective cavity mode frequency, and $\omega_0$ is the frequency splitting of the atomic levels. $\hat{J}_{\pm,z}$ are the collective angular momentum operators given by $\hat{J}_{\pm,z} = \sum_{\nu=1}^{N}\hat{\sigma}_{\pm,z}^{(\nu)},$ where $\hat{\sigma}_{\pm,z}^{(\nu)}$ are the spin-$\frac{1}{2}$ Pauli operators for the $\nu$th atom. We consider the case of indistinguishable atoms, with maximal total angular momentum $J=N/2$.
	
	The dominant source of damping in the experiments of \cite{zhiqiang_nonequilibrium_2017} is cavity decay. To this end, we model dissipation with the quantum master equation
	\begin{equation}\label{eqn:master}
	\frac{d\hat{\rho}}{dt} = -i[\hat{H},\hat{\rho}] + \kappa\left( 2\hat{a}\hat{\rho}\hat{a}^{\dagger} - \hat{a}^{\dagger}\hat{a}\hat{\rho} - \hat{\rho}\hat{a}^{\dagger}\hat{a} \right),
	\end{equation}
	where $\hat{\rho}$ is the reduced density operator of the atoms and cavity mode, and $\kappa$ is the cavity field decay rate.
	
	This model features a parity symmetry with the operator $\hat{\Pi} = \exp[{i\pi( \hat{a}^{\dagger}\hat{a} + \hat{J}_{z} + J )}]$, which acts on the system operators as $\hat{\Pi}^{\dagger}\hat{a}\hat{\Pi}=-\hat{a}$, $\hat{\Pi}^{\dagger}\hat{J}_{-}\hat{\Pi}=-\hat{J}_{-}$, and $\hat{\Pi}^{\dagger}\hat{J}_{z}\hat{\Pi}=\hat{J}_{z}$. Both the Hamiltonian and the Lindbladian cavity-decay superoperator are invariant under this transformation, giving rise to the system's parity symmetry, which is broken by the superradiant phase transition.
	
	
	\textit{Steady state analysis.}---We define the variables 
	\begin{equation*}
 \alpha = \braket{\hat{a}}/\sqrt{N} \in \mathbb{C},\ \beta = \langle \hat{J}_{-} \rangle/N  \in \mathbb{C},\ \gamma = \langle \hat{J}_{z} \rangle/N \in\mathbb{R}
	\end{equation*}
and, in the semiclassical limit $N\rightarrow\infty$, factorize operator expectations, i.e., $\langle \hat{a}\hat{J}_{+} \rangle \approx \langle \hat{a} \rangle \langle \hat{J}_{+} \rangle$. With this approximation, the master equation (\ref{eqn:master}) leads to a system of nonlinear differential equations that determine the generalized Dicke model's semiclassical dynamics:
	\begin{subequations}\label{eqn:ODEs}
		\begin{align}
		\frac{d\alpha}{dt} &= -\kappa\alpha - i\omega\alpha - i\lambda_{-}\beta - i\lambda_{+}\beta^*, \\
		\frac{d\beta}{dt} &= -i\omega_0 \beta + 2i\lambda_{-}\alpha\gamma + 2i\lambda_{+}\alpha^{*}\gamma, \\
		\frac{d\gamma}{dt} &= i\lambda_{-}\left( \alpha^{*}\beta - \alpha\beta^{*} \right) + i\lambda_{+}\left( \alpha\beta - \alpha^{*}\beta^{*} \right).
		\end{align}
	\end{subequations}
	A comprehensive analysis of the dynamics described by these equations is undertaken in \cite{stitely_nonlinear_2020}. We first turn our attention to comparing the steady states of the semiclassical model (\ref{eqn:ODEs}) and the fully quantum model given by the master equation (\ref{eqn:master}).
	
	
	\begin{figure}[t!]
		\centering
		\includegraphics[width=8.6cm]{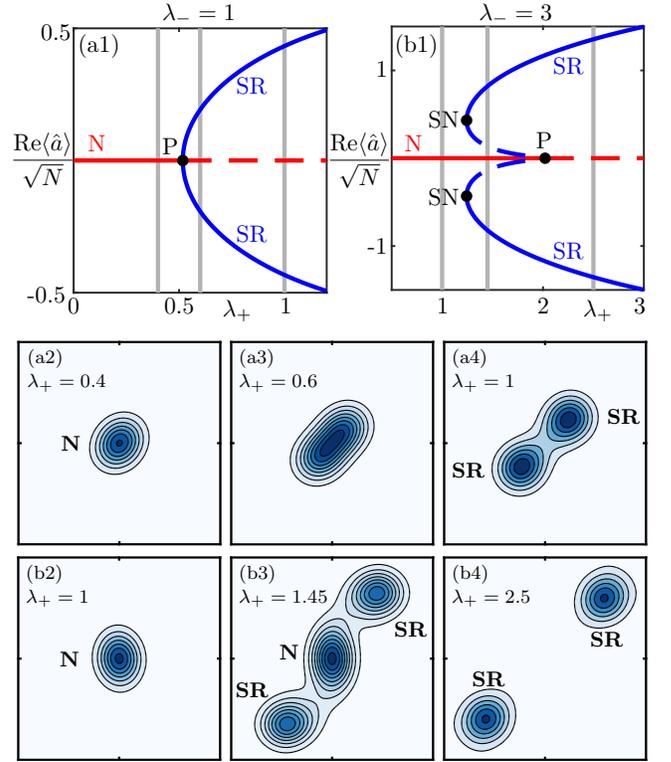}
		\caption{\label{fig:QFunc}One- and two-stage superradiant phase transitions in the unbalanced Dicke model for $\lambda_{-} = 1$ and $\lambda_{-} = 3$. Panels (a1) and (b1) are the two corresponding bifurcation diagrams of the semiclassical model, showing the normal phase (red) and superradiant equilibria (blue); here unstable equilibria are shown dashed.
		Below are steady-state Husimi Q-functions evaluated along the single stage transition (a2)--(a4) and the two-stage transition (b2)--(b4), plotted on the $\chi$-plane in a square from -6.5 to 6.5 in the real and complex axes. Darker color indicates a larger value of $Q(\chi)$. Here $\kappa = \omega = \omega_0 = 1$ and $N=8$.}
	\end{figure}
	
	The superradiant phase is defined by a non-zero photon number in the steady state, while the vacuum state is referred to as the normal phase. In the semiclassical regime, the superradiant phase transition in the case of balanced coupling ($\lambda_{-}=\lambda_{+}$) is described by a pitchfork bifurcation, where a stable equilibrium point bifurcates into two stable equilibria representing the superradiant states and an unstable equilibrium representing the normal phase.
	
	For unbalanced coupling ($\lambda_{-}\neq\lambda_{+}$), this situation is preserved for some constant values of $\lambda_{-}$ (increasing $\lambda_{+}$) with the transition now taking place at a new threshold value for $\lambda_{+}$ (see \cite{stitely_nonlinear_2020} for details). Figure \ref{fig:QFunc}(a1) shows a bifurcation diagram for this transition; here we plot the equilibria of system (\ref{eqn:ODEs}) as $\lambda_{+}$ is varied for $\lambda_{-} = 1$. The curves N and SR are the normal and superradiant equilibria, respectively; unstable equilibria are shown dashed. The superradiant curves emerge from the normal phase in a pitchfork bifurcation (P), which turns the normal phase unstable.
	
	In addition to this standard case, for larger values of $\lambda_{-}$, the superradiant phase transition is split into \emph{two stages}, with first the onset of superradiance and a subsequent disappearance of the normal phase. This is illustrated in a bifurcation diagram in Fig.~\ref{fig:QFunc}(b1). Starting from the normal phase, as $\lambda_{+}$ increases the system undergoes two simultaneous saddle-node bifurcations (SN), which create a pair of stable and unstable superradiant equilibria each. This is a first-order phase transition, responsible for the creation of large amplitude superradiant states. The two unstable equilibria then disappear by colliding with the normal phase equilibrium point in a pitchfork bifurcation (P), turning the normal phase unstable. This configuration of the two bifurcations creates a region of multistability, between the saddle-node and pitchfork bifurcations, where there are coexisting normal and superradiant phases \cite{soriente_dissipation-induced_2018,keeling_collective_2010}. Moreover, there is a critical point in the $(\lambda_{-},\lambda_{+})$-plane from which multistability emerges, where the saddle-node and pitchfork bifurcations coincide at $\lambda_{\pm}^*=\sqrt{\omega_0(\zeta\mp\omega\sqrt{\zeta})/2\omega}$, where $\zeta=\kappa^2+\omega^2$ (see \cite{stitely_nonlinear_2020} for details).
	
	
	To compare the semiclassical description to the quantum phase transition for finite $N$, we solve the master equation (\ref{eqn:master}) in the steady state to obtain the steady-state density operator $\hat{\rho}_{\mathrm{ss}}$. The quantum phase transition is visualized with the Husimi Q-function of the light field, obtained by tracing over the atomic states, $Q(\chi) = \braket{\chi|\Tr_{A}(\hat{\rho}_{\mathrm{ss}})|\chi}/\pi$, where $\ket{\chi}$ is a coherent state and $\chi\in\mathbb{C}$.
	
	Figure.~\ref{fig:QFunc}(a2)--(a4) shows the Husimi Q-function for three values of $\lambda_{+}$ along the single-stage transition. This gives the quantum analogue of the stable equilibria of the semiclassical system, which are now represented as peaks of a distribution rather than points in phase space that represent the mean-field behavior. Initially, the system is in the normal phase [Fig.~\ref{fig:QFunc}(a2)]. As the coupling strength $\lambda_{+}$ is increased, the peak splits [Fig.~\ref{fig:QFunc}(a3)] as two superradiant states emerge from the normal phase [Fig.~\ref{fig:QFunc}(a4)]. Note that, in contrast to the semiclassical transition, due to finite size effects, the quantum transition has no clearly defined critical point.
	
	The quantum analogue of the two-stage superradiant phase transition is shown in Fig.~\ref{fig:QFunc}(b2)--(b4). Again, the system begins in the normal phase [Fig.~\ref{fig:QFunc}(b2)]. As $\lambda_{+}$ increases, two superradiant states spontaneously emerge and the system becomes multistable, with coexisting normal and superradiant phases, leading to a three-peaked Q-function [Fig.~\ref{fig:QFunc}(b3)]. Unlike the above transition, this first-order transition is less affected by finite size effects, and therefore has a more clearly defined critical point. The emergence of these states is the quantum analogue of the pair of saddle-node bifurcations in Fig.~\ref{fig:QFunc}(b1). As $\lambda_{+}$ increases further across the second stage of the transition, corresponding to the pitchfork bifurcation, the normal phase peak disappears, and the system is entirely superradiant [Fig.~\ref{fig:QFunc}(b4)]. It is quite remarkable that the semiclassical model can give such insight into the quantum phase transitions for such small numbers of atoms as $N=8$ as in Fig. ~\ref{fig:QFunc}.
	
\begin{figure}[t]
		\centering
		\includegraphics[width=8.6cm]{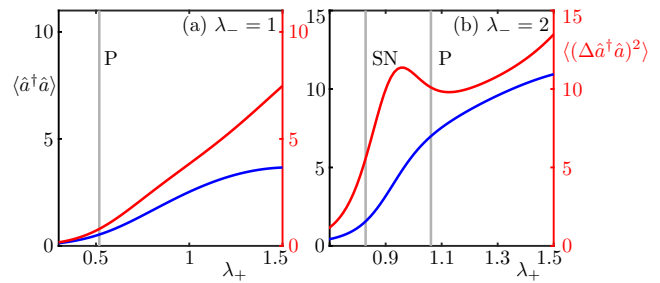}
		\caption{\label{fig:expect}Steady state photon number expectation $\langle \hat{a}^{\dagger}\hat{a} \rangle$ (blue) and variance $\langle (\Delta \hat{a}^{\dagger}\hat{a})^2 \rangle$ (red) as $\lambda_{+}$ is varied. Panel (a) shows the single-stage superradiant phase transition when $\lambda_{-}=1$. Panel (b) shows the two-stage superradiant phase transition when $\lambda_{-}=2$. The bifurcations P and SN  of the semiclassical model are indicated as gray vertical lines. Here $\kappa = \omega = \omega_0 = 1$ and $N=8$.}
	\end{figure}

	As Fig.~\ref{fig:expect} shows, evidence of a two-stage superradiant phase transition is also reflected in the steady-state photon number and photon number variance, again for only $N=8$ atoms. The single-stage case is illustrated in Fig.~\ref{fig:expect}(a), where both the photon number and variance begin to increase near the pitchfork bifurcation (P) of the semiclassical model, after which they grow when the system becomes superradiant. In the two-stage case, on the other hand, the photon number grows very quickly in the $\lambda_{+}$-range between the pitchfork and saddle-node bifurcations [Fig.~\ref{fig:expect}(b)]. This sharp growth is attributed to the spontaneous emergence of the superradiant phase and the onset of multistability. As $\lambda_{+}$ is further increased, the normal phase disappears and the photon number grows more slowly once the system is entirely in the superradiant phase.
	
	The two-stage onset of superradiance appears even more clearly in the variance in Fig.~\ref{fig:expect}(b). In particular, the variance begins to increase rapidly near the saddle-node bifurcation (SN). This is because, in this regime of multistability, the normal and superradiant phases are in a statistical mixture. When the superradiant steady states first emerge, their influence is small and the system is dominated by the normal phase. As $\lambda_{+}$ increases through the region of multistability, the normal and superradiant phases compete until they gain equal influence on the global state. Here the variance reaches a local maximum. After this point the superradiant phase becomes dominant as the normal phase disappears, and the variance decreases. It then reaches a local minimum, where the decrease in the variance due to the disappearance of the normal phase is balanced by the increase due to the superradiant states increasing in amplitude. Notice that since $\langle (\Delta\hat{a}^{\dagger}\hat{a})^2 \rangle > \langle \hat{a}^{\dagger}\hat{a} \rangle$, the statistics of the field are always super-Poissonian.
	
	
	\textit{Quantum dynamics}.---The emergence of multistability in the steady state has a profound effect on the quantum dynamics of the system. In the multistable region, various stable states can be accessed by the system through quantum tunnelling. Due to the system's parity symmetry, there exist a pair of stable superradiant states, related by the substitution $\hat{a}\rightarrow -\hat{a}$, $\hat{J}_{-}\rightarrow -\hat{J}_{-}$. The two states are indistinguishable in the photon number as $\hat{a}^{\dagger}\hat{a}$ is invariant under the parity transformation. However, the two superradiant states can be distinguished with the use of a stochastic Schr\"{o}dinger equation \cite{carmichael_open_1993,carmichael_statistical_2007} to simulate heterodyne detection of the quadrature operator
	\begin{equation}
	\hat{X}_{-\pi/4} = \frac{1}{\sqrt{2}}\left( \hat{a}e^{-i\pi/4} + \hat{a}^{\dagger} e^{i\pi/4}\right).
	\end{equation}
	This method allows for an emulation of a phase sensitive detection scheme, so that the two superradiant states may be distinguished; see supplemental material \footnote{See supplemental material below for a brief discussion of quantum trajectories and the heterodyne detection stochastic Schr{\"o}dinger equation.} for details.

	\begin{figure}[t]
		\centering
		\includegraphics[width=8.6cm]{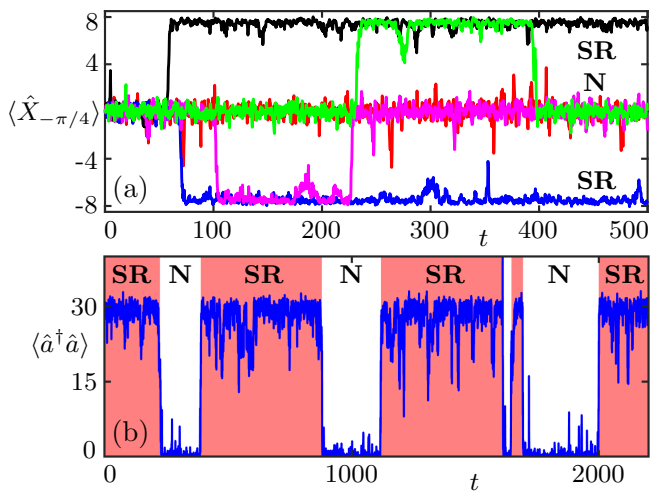}
		\caption{\label{fig:switching}Quantum jumps between normal and superradiant phases. Panel (a) shows five illustrative stochastic realizations monitored via a heterodyne detection scheme. Panel (b) shows superradiant switching in the photon number of a single realization, also monitoring the $\hat{X}_{-\pi/4}$ quadrature. Here $\kappa = \omega = \omega_0 = 1$, $\lambda_{-} = 3$, $\lambda_{+} = 1.45$, and $N=15$.}
	\end{figure}

	 Figure~\ref{fig:switching}(a) shows five realizations with identical initial conditions, namely, ground states for the atoms and vacuum state for the light field. The coexistent normal and superradiant states can be seen clearly as trajectories are sustained around three distinct attractors: the two superradiant states with $\langle \hat{X}_{-\pi/4} \rangle\neq 0$ and the normal phase with $\langle \hat{X}_{-\pi/4} \rangle\approx 0$. We also observe instances of tunnelling between phases, where realizations initially in the normal phase spontaneously jump to the superradiant phase, driven by quantum fluctuations. The presence of a stable normal phase disrupts direct tunnelling of trajectories between the two superradiant states, because to do so trajectories would need to pass through a region of phase space which is now attracting, hence, pulling trajectories into the normal phase.
	
	On a longer timescale, the system continually switches between the normal and superradiant phase as quantum jumps drive the system between stable states. As Fig.~\ref{fig:switching}(b) shows, this creates dramatic changes in the photon number as trajectories move between the normal and superradiant states. As the classical limit is approached, the relative strength of quantum fluctuations reduces and the switching rate gradually decreases. In the limit $N\rightarrow\infty$, the system becomes classically deterministic and the final state is fully determined by the initial condition. 

	
	\textit{Quantum hysteresis.}---Systems with multistable states generically feature history dependence, known as \emph{hysteresis}, when parameters are swept. In this situation, during a parameter sweep certain equilibria can only be accessed when increasing or decreasing a parameter. A hysteresis loop exists in the semiclassical model due to the multistable regime between the saddle-node and pitchfork bifurcations. If a slow, adiabatic sweep is performed from the normal phase for increasing $\lambda_{+}$, the system remains in the normal phase until the pitchfork bifurcation (P) is reached and then it quickly transitions to the superradiant phase. If instead $\lambda_{+}$ is decreased from the superradiant phase, the system tracks along the superradiant branch all the way to the saddle-node bifurcation (SN); the system then makes a fast transition to the normal phase.
	
	\begin{figure}[t]
		\centering
		\includegraphics[width=0.4\textwidth]{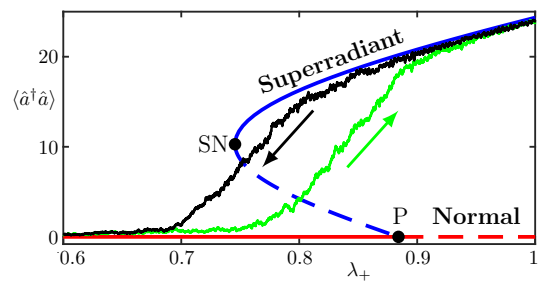}
		\caption{\label{fig:hysteresis}Quantum hysteresis in the photon number $\langle \hat{a}^{\dagger}\hat{a} \rangle$, monitored via photocounting. Shown is the ensemble average of fifty quantum trajectories when $\lambda_{+}$ is increased (green), and when $\lambda_{+}$ is decreased (black); also shown is the semiclassical bifurcation diagram with normal phase equilibria (red) and the superradiant equilibria (blue), where unstable equilibria are shown dashed. Here $\kappa = \omega = \omega_0 = 1$, $\lambda_{-}=1.8$, $N=30$, and $T_{\mathrm{ramp}}=500/\kappa$.}
	\end{figure}
	
	The semiclassical hysteresis loop has a quantum mechanical counterpart. Figure~\ref{fig:hysteresis} shows the ensemble average over a series of fifty (photon counting) quantum trajectories subject to a slow, linear time-dependence on $\lambda_{+}$, either increasing or decreasing, overlaid atop the semiclassical bifurcation diagram. We used a ramp time of $T_{\mathrm{ramp}}=500/\kappa$ for $N=30$ atoms, which was chosen to be considerably slower than the equilibration time, to ensure the adiabatic tracking of the quantum states. The ramp time is also considerably shorter than the switching time, otherwise averaging over many switches will reproduce the master equation result. The quantum hysteresis loop in Fig.~\ref{fig:hysteresis} is in good agreement with the predictions from the semiclassical model. However, there are some notable differences. In the upward sweep of $\lambda_{+}$, the photon number begins to increase before the pitchfork bifurcation due to quantum tunnelling. Similarly, in the downward sweep, the photon number begins to decrease before the saddle-node bifurcation, again due to quantum tunnelling, and remains non-zero for a short time even after the bifurcation. Indeed, the right hand side of system (\ref{eqn:ODEs}) is near-zero in the region where the equilibria just disappeared in the saddle-node bifurcation (SN), this produces a ``quasistable" superradiant area in phase space, also referred to as a \emph{ghost} \cite{strogatz_nonlinear_2015}, that can be tracked during the downward sweep of the semiclassical model. With the addition of quantum fluctuations, the quasistable region can be accessed by tunnelling, so trajectories can linger there for some periods of time during the downward sweep of $\lambda_{+}$, causing the photon number to remain non-zero after the SN point. Contrast this with the pitchfork bifurcation (P), where the stable normal phase not only disappears, but is replaced by an unstable state that \emph{actively repels} trajectories to drive them to the superradiant states, thus, not creating a possibility for transient quasistability in the upward sweep.
	
	
	\begin{figure}[t]
		\centering
		\includegraphics[width=8.6cm]{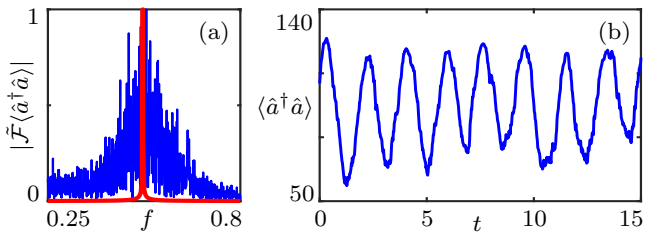}
		\caption{\label{fig:oscillations}Superradiant oscillations in the quantum regime. Panel (a) shows the (normalized) Fourier spectra of a semiclassical (red) and quantum (blue) photon number trajectory, where $\tilde{\mathcal{F}}\{\cdot\} = \mathcal{F}\{\cdot\}/\max\mathcal{F}\{\cdot\}$. Panel (b) shows the temporal trace of the photon number oscillations of a quantum trajectory. Here $\kappa = \omega = \omega_0 = 1$, $\lambda_{-} = 1.5$, $\lambda_{+} = 1.8$, $N = 200$, and the total sample time is 2000.}
	\end{figure}

	\textit{Superradiant oscillations.}---A major result of the experiments in \cite{zhiqiang_nonequilibrium_2017} is the observation of an oscillatory superradiant phase, identified in a short-lived (due to dephasing) cavity output signal. In the semiclassical description, these oscillations emerge from a Hopf bifurcation of superradiant equilibria (see \cite{stitely_nonlinear_2020} for details). Figure \ref{fig:oscillations} shows that such oscillations can also be identified in quantum simulations of (photocounting) quantum trajectories. Compared to equilibria, the clear detection of oscillations requires a significantly greater reduction of quantum fluctuations, so we use a higher atom number of $N=200$. Figure \ref{fig:oscillations}(a) shows that the Fourier spectrum of the quantum trajectories, while significantly broadened, peaks at the single main frequency of the semiclassical oscillation. The spectrum encodes the oscillatory property of the quantum trajectory in Fig. \ref{fig:oscillations}(b), subject to quantum fluctuations. Importantly, for only $N=200$ atoms these fluctuations are small enough to reveal the oscillations of the semiclassical limit. 
	
	
	\textit{Conclusions and outlook.}---We have studied the quantum signatures of nonlinear dynamics in a generalized, open Dicke model. Our results provide fundamental insight into the quantum--classical transition by explaining how features of classical nonlinear dynamics arise from the quantum world. Specifically, we have shown that the presence of quantum fluctuations during a two-stage transition to superradiance causes spontaneous switching between the normal and superradiant phases, and that this situation leads to the formation of a quantum hysteresis loop when parameters are swept adiabatically. Given the success of recent experiments, our analysis suggests that the multistable region may be of considerable interest to probe experimentally. Moreover, we have demonstrated that superradiant oscillations can be identified for atom numbers several orders of magnitude smaller than those in current experiments. This suggests that it should be possible to access experimentally also the quantum signatures of more complex nonlinear phenomena observed in the semiclassical analysis \cite{stitely_nonlinear_2020}; including other types of periodic oscillations, as well as localized and non-localized chaotic atttractors. Overall, our results demonstrate that insights offered by semiclassical models extend further into the quantum world than previously thought.

	\bibliography{SMGKP_Switching.bib}

\begin{thebibliography}{10}

\bibitem{loudon_quantum_1983}
R.~Loudon, {\em The quantum theory of light}.
\newblock Clarendon Press, 3rd~ed., Jan. 1983.

\bibitem{scully_quantum_1997}
M.~O. Scully and M.~S. Zubairy, {\em Quantum {Optics}}.
\newblock Cambridge University Press, 1997.

\bibitem{gross_superradiance:_1982}
M.~Gross and S.~Haroche, ``Superradiance: {An} essay on the theory of
  collective spontaneous emission,'' {\em Physics Reports}, vol.~93,
  pp.~301--396, Dec. 1982.

\bibitem{dicke_coherence_1954}
R.~H. Dicke, ``Coherence in {Spontaneous} {Radiation} {Processes},'' {\em Phys.
  Rev.}, vol.~93, pp.~99--110, Jan. 1954.

\bibitem{hepp_equilibrium_1973}
K.~Hepp and E.~H. Lieb, ``Equilibrium {Statistical} {Mechanics} of {Matter}
  {Interacting} with the {Quantized} {Radiation} {Field},'' {\em Phys. Rev. A},
  vol.~8, pp.~2517--2525, Nov. 1973.

\bibitem{hepp_superradiant_1973}
K.~Hepp and E.~H. Lieb, ``On the {Superradiant} {Phase} {Transition} for
  {Molecules} in a {Quantized} {Radiation} {Field}: the {Dicke} {Maser}
  {Model},'' {\em Annals of Physics}, vol.~76, no.~2, pp.~360 -- 404, 1973.

\bibitem{wang_phase_1973}
Y.~K. Wang and F.~T. Hioe, ``Phase {Transition} in the {Dicke} {Model} of
  {Superradiance},'' {\em Phys. Rev. A}, vol.~7, pp.~831--836, Mar. 1973.

\bibitem{kirton_introduction_2019}
P.~Kirton, M.~M. Roses, J.~Keeling, and E.~G. Dalla~Torre, ``Introduction to
  the {Dicke} {Model}: {From} {Equilibrium} to {Nonequilibrium}, and {Vice}
  {Versa},'' {\em Advanced Quantum Technologies}, vol.~2, p.~1970013, Feb.
  2019.

\bibitem{baumann_dicke_2010}
K.~Baumann, C.~Guerlin, F.~Brennecke, and T.~Esslinger, ``Dicke quantum phase
  transition with a superfluid gas in an optical cavity,'' {\em Nature},
  vol.~464, pp.~1301--1306, July 2010.

\bibitem{baumann_exploring_2011}
K.~Baumann, R.~Mottl, F.~Brennecke, and T.~Esslinger, ``Exploring {Symmetry}
  {Breaking} at the {Dicke} {Quantum} {Phase} {Transition},'' {\em Phys. Rev.
  Lett.}, vol.~107, p.~140402, Sept. 2011.

\bibitem{hamner_dicke-type_2014}
C.~Hamner, C.~Qu, Y.~Zhang, J.~Chang, M.~Gong, C.~Zhang, and P.~Engels,
  ``Dicke-type phase transition in a spin-orbit-coupled {Bose}-{Einstein}
  condensate,'' {\em Nat Commun}, vol.~5, p.~4023, June 2014.

\bibitem{safavi-naini_verification_2018}
A.~Safavi-Naini, R.~J. Lewis-Swan, J.~G. Bohnet, M.~G\"arttner, K.~A. Gilmore,
  J.~E. Jordan, J.~Cohn, J.~K. Freericks, A.~M. Rey, and J.~J. Bollinger,
  ``Verification of a many-ion simulator of the dicke model through slow
  quenches across a phase transition,'' {\em Phys. Rev. Lett.}, vol.~121,
  p.~040503, Jul 2018.

\bibitem{dimer_proposed_2007}
F.~Dimer, B.~Estienne, A.~S. Parkins, and H.~J. Carmichael, ``Proposed
  realization of the {Dicke}-model quantum phase transition in an optical
  cavity {QED} system,'' {\em Phys. Rev. A}, vol.~75, p.~013804, Jan. 2007.

\bibitem{zhiqiang_nonequilibrium_2017}
Z.~Zhiqiang, C.~H. Lee, R.~Kumar, K.~J. Arnold, S.~J. Masson, A.~S. Parkins,
  and M.~D. Barrett, ``{Nonequilibrium} phase transition in a spin-1 {Dicke}
  model,'' {\em Optica}, vol.~4, pp.~424--429, Apr 2017.

\bibitem{zhang_dicke_2018}
Z.~Zhang, C.~H. Lee, R.~Kumar, K.~J. Arnold, S.~J. Masson, A.~L. Grimsmo, A.~S.
  Parkins, and M.~D. Barrett, ``Dicke-model simulation via cavity-assisted
  raman transitions,'' {\em Phys. Rev. A}, vol.~97, p.~043858, Apr 2018.

\bibitem{stitely_nonlinear_2020}
K.~C. Stitely, A.~Giraldo, B.~Krauskopf, and S.~Parkins, ``Nonlinear
  semiclassical dynamics of the unbalanced, open dicke model,'' {\em Phys. Rev.
  Research}, vol.~2, p.~033131, Jul 2020.

\bibitem{gelhausen_dissipative_2018}
J.~Gelhausen and M.~Buchhold, ``Dissipative dicke model with collective atomic
  decay: Bistability, noise-driven activation, and the nonthermal first-order
  superradiance transition,'' {\em Phys. Rev. A}, vol.~97, p.~023807, Feb 2018.

\bibitem{soriente_dissipation-induced_2018}
M.~Soriente, T.~Donner, R.~Chitra, and O.~Zilberberg, ``Dissipation-induced
  anomalous multicritical phenomena,'' {\em Phys. Rev. Lett.}, vol.~120,
  p.~183603, May 2018.

\bibitem{keeling_collective_2010}
J.~Keeling, M.~J. Bhaseen, and B.~D. Simons, ``Collective dynamics of
  bose-einstein condensates in optical cavities,'' {\em Phys. Rev. Lett.},
  vol.~105, p.~043001, Jul 2010.

\bibitem{carmichael_open_1993}
H.~J. Carmichael, {\em An {Open} {Systems} {Approach} to {Quantum} {Optics}}.
\newblock Springer Science \& Business Media, May 1993.

\bibitem{carmichael_statistical_2007}
H.~J. Carmichael, {\em Statistical {Methods} in {Quantum} {Optics} 2:
  {Non}--{Classical} {Fields}}.
\newblock Theoretical and {Mathematical} {Physics}, Springer, Jan. 2007.

\bibitem{Note1}
See supplemental material below for a brief discussion of quantum trajectories
  and the heterodyne detection stochastic Schr{\"o}dinger equation.

\bibitem{strogatz_nonlinear_2015}
S.~Strogatz, {\em Nonlinear Dynamics and Chaos: With Applications to Physics,
  Biology, Chemistry, and Engineering}, vol.~2.
\newblock Westview Press, 2015.

\end{thebibliography}


\begin{thebibliography}{5}%
\makeatletter
\providecommand \@ifxundefined [1]{%
 \@ifx{#1\undefined}
}%
\providecommand \@ifnum [1]{%
 \ifnum #1\expandafter \@firstoftwo
 \else \expandafter \@secondoftwo
 \fi
}%
\providecommand \@ifx [1]{%
 \ifx #1\expandafter \@firstoftwo
 \else \expandafter \@secondoftwo
 \fi
}%
\providecommand \natexlab [1]{#1}%
\providecommand \enquote  [1]{``#1''}%
\providecommand \bibnamefont  [1]{#1}%
\providecommand \bibfnamefont [1]{#1}%
\providecommand \citenamefont [1]{#1}%
\providecommand \href@noop [0]{\@secondoftwo}%
\providecommand \href [0]{\begingroup \@sanitize@url \@href}%
\providecommand \@href[1]{\@@startlink{#1}\@@href}%
\providecommand \@@href[1]{\endgroup#1\@@endlink}%
\providecommand \@sanitize@url [0]{\catcode `\\12\catcode `\$12\catcode
  `\&12\catcode `\#12\catcode `\^12\catcode `\_12\catcode `\%12\relax}%
\providecommand \@@startlink[1]{}%
\providecommand \@@endlink[0]{}%
\providecommand \url  [0]{\begingroup\@sanitize@url \@url }%
\providecommand \@url [1]{\endgroup\@href {#1}{\urlprefix }}%
\providecommand \urlprefix  [0]{URL }%
\providecommand \Eprint [0]{\href }%
\providecommand \doibase [0]{http://dx.doi.org/}%
\providecommand \selectlanguage [0]{\@gobble}%
\providecommand \bibinfo  [0]{\@secondoftwo}%
\providecommand \bibfield  [0]{\@secondoftwo}%
\providecommand \translation [1]{[#1]}%
\providecommand \BibitemOpen [0]{}%
\providecommand \bibitemStop [0]{}%
\providecommand \bibitemNoStop [0]{.\EOS\space}%
\providecommand \EOS [0]{\spacefactor3000\relax}%
\providecommand \BibitemShut  [1]{\csname bibitem#1\endcsname}%
\let\auto@bib@innerbib\@empty
\bibitem [{\citenamefont {Carmichael}(2007)}]{carmichael_statistical_2007}%
  \BibitemOpen
  \bibfield  {author} {\bibinfo {author} {\bibfnamefont {H.~J.}\ \bibnamefont
  {Carmichael}},\ }\href {\doibase 10.1007/978-3-540-71320-3} {\emph {\bibinfo
  {title} {Statistical {Methods} in {Quantum} {Optics} 2: {Non}--{Classical}
  {Fields}}}},\ Theoretical and {Mathematical} {Physics}\ (\bibinfo
  {publisher} {Springer},\ \bibinfo {year} {2007})\BibitemShut {NoStop}%
\bibitem [{\citenamefont {Daley}(2014)}]{daley_manyBodyTrajectories_2014}%
  \BibitemOpen
  \bibfield  {author} {\bibinfo {author} {\bibfnamefont {A.~J.}\ \bibnamefont
  {Daley}},\ }\href {\doibase 10.1080/00018732.2014.933502} {\bibfield
  {journal} {\bibinfo  {journal} {Advances in Physics}\ }\textbf {\bibinfo
  {volume} {63}},\ \bibinfo {pages} {77} (\bibinfo {year} {2014})}\BibitemShut
  {NoStop}%
\bibitem [{\citenamefont {Carmichael}(1993)}]{carmichael_open_1993}%
  \BibitemOpen
  \bibfield  {author} {\bibinfo {author} {\bibfnamefont {H.~J.}\ \bibnamefont
  {Carmichael}},\ }\href@noop {} {\emph {\bibinfo {title} {An {Open} {Systems}
  {Approach} to {Quantum} {Optics}}}}\ (\bibinfo  {publisher} {Springer Science
  \& Business Media},\ \bibinfo {year} {1993})\BibitemShut {NoStop}%
\bibitem [{\citenamefont {Wiseman}\ and\ \citenamefont
  {Milburn}(1993{\natexlab{a}})}]{wiseman_quantum_1993}%
  \BibitemOpen
  \bibfield  {author} {\bibinfo {author} {\bibfnamefont {H.~M.}\ \bibnamefont
  {Wiseman}}\ and\ \bibinfo {author} {\bibfnamefont {G.~J.}\ \bibnamefont
  {Milburn}},\ }\href {\doibase 10.1103/PhysRevA.47.642} {\bibfield  {journal}
  {\bibinfo  {journal} {Phys. Rev. A}\ }\textbf {\bibinfo {volume} {47}},\
  \bibinfo {pages} {642} (\bibinfo {year} {1993}{\natexlab{a}})}\BibitemShut
  {NoStop}%
\bibitem [{\citenamefont {Wiseman}\ and\ \citenamefont
  {Milburn}(1993{\natexlab{b}})}]{wiseman_interpretation_1993}%
  \BibitemOpen
  \bibfield  {author} {\bibinfo {author} {\bibfnamefont {H.~M.}\ \bibnamefont
  {Wiseman}}\ and\ \bibinfo {author} {\bibfnamefont {G.~J.}\ \bibnamefont
  {Milburn}},\ }\href {\doibase 10.1103/PhysRevA.47.1652} {\bibfield  {journal}
  {\bibinfo  {journal} {Phys. Rev. A}\ }\textbf {\bibinfo {volume} {47}},\
  \bibinfo {pages} {1652} (\bibinfo {year} {1993}{\natexlab{b}})}\BibitemShut
  {NoStop}%
\end{thebibliography}%
	
\end{document}


\title{Supplemental Material for ``Superradiant Switching, Quantum Hysteresis, and Oscillations in a Generalized Dicke Model"}
	
\author{Kevin C. Stitely}
\affiliation{Dodd-Walls Centre for Photonic and Quantum Technologies, New Zealand}
\affiliation{Department of Mathematics, University of Auckland, Auckland 1010, New Zealand}
\affiliation{Department of Physics, University of Auckland, Auckland 1010, New Zealand}
	
\author{Stuart J. Masson}
\affiliation{Department of Physics, Columbia University, New York, New York 10027, USA}
	
\author{Andrus Giraldo}
\affiliation{Dodd-Walls Centre for Photonic and Quantum Technologies, New Zealand}
\affiliation{Department of Mathematics, University of Auckland, Auckland 1010, New Zealand}
	
\author{Bernd Krauskopf}
\affiliation{Dodd-Walls Centre for Photonic and Quantum Technologies, New Zealand}
\affiliation{Department of Mathematics, University of Auckland, Auckland 1010, New Zealand}
	
\author{Scott Parkins}
\affiliation{Dodd-Walls Centre for Photonic and Quantum Technologies, New Zealand}
\affiliation{Department of Physics, University of Auckland, Auckland 1010, New Zealand}
	
	
\maketitle

In these supplementary materials we provide details on quantum trajectories and stochastic Schr\"{o}dinger equations used to give single realizations of an open quantum system. The stochastic Schr\"{o}dinger approach is used to access phase information of the system by simulating interference of the output of the cavity with a strong laser source with a beamsplitter.

\section{Quantum Trajectory Theory: Photocounting}
Quantum trajectory theory \cite{carmichael_statistical_2007,daley_manyBodyTrajectories_2014,carmichael_open_1993} provides an avenue for ``unravelling" the master equation
\begin{equation}\label{eqn:MasterEqn}
\frac{d\hat{\rho}}{dt} = -i[\hat{H},\hat{\rho}] + \kappa\left( 2\hat{a}\hat{\rho}\hat{a}^{\dagger} - \hat{a}^{\dagger}\hat{a}\hat{\rho} - \hat{\rho}\hat{a}^{\dagger}\hat{a} \right)
\end{equation}
in terms of a stochastic average over an ensemble of individual realizations of the dynamics as a set of pure states. The quantum trajectory approach splits the master equation into two contributions: a continuous non-unitary evolution and randomly applied discrete quantum jumps. The former involves evolution described by a Schr\"{o}dinger equation with the non-Hermitian Hamiltonian
\begin{equation}\label{eqn:nonHermitianHamiltonian}
\hat{H}_{\mathrm{eff}} = \hat{H} - i\kappa\hat{a}^{\dagger}\hat{a}.
\end{equation}
During the course of a single realization, the quantum state is time-evolved under the non-Hermitian Hamiltonian $\hat{H}_{\mathrm{eff}}$, interrupted by quantum jumps. Starting with an initial pure state $\ket{\psi_0}$, at each step $j=0,\dots,n$ with a time increment $\delta t$ the probability of the occurence of a quantum jump is $p_j = 2\kappa\braket{\psi_j|\hat{a}^{\dagger}\hat{a}|\psi_j}\delta t$. Written as an algorithm describing a single quantum trajectory under photocounting at each step $j$, the quantum state $\ket{\psi_{j+1}}$ is computed as follows:
\begin{itemize}[itemindent=5\parindent,itemsep=-3mm]
	\item[\textbf{Step 1}:] Advance simulation time $t_{j+1} = t_{j} + \delta t$. \\
	\item[\textbf{Step 2}:] Calculate the probability of a quantum jump $p_{j} = 2\kappa\braket{\psi_{j}|\hat{a}^{\dagger}\hat{a}|\psi_{j}}\delta t$. \\
	\item[\textbf{Step 3}:] Sample a random number $r_{j}\in [0,1]$ from a uniform probability distribution. \\
	\item[\textbf{Step 4}:] If $p_j>r_j$ (quantum jump):
		\begin{equation}
		\ket{\psi_{j+1}} = \frac{\hat{a}\ket{\psi_j}}{\sqrt{\braket{\psi_j|\hat{a}^{\dagger}\hat{a}|\psi_j}}}.
		\end{equation} \\
	\item[\textbf{Step 5}:] Else (evolution by non-Hermitian Hamiltonian):
		\begin{equation}
		\ket{\psi_{j+1}} = \frac{e^{-i\hat{H}_{\mathrm{eff}}\delta t}\ket{\psi_j}}{\sqrt{\braket{\psi_j | e^{i\hat{H}_{\mathrm{eff}}^{\dagger}\delta t}e^{-i\hat{H}_{\mathrm{eff}}\delta t} | \psi_j}}}.
		\end{equation}
\end{itemize}
Under these rules for the system's time-evolution, quantum trajectories are consistent with the master equation (\ref{eqn:MasterEqn}) in that the ensemble average of the expectation values of operators over many realizations will recreate the expectation values obtained by tracing over the density matrix solution to the master equation.

\section{Stochastic Schr\"{o}dinger equations: Homodyne and Heterodyne Detection}
In the quantum trajectories framework the choice of how to unravel the master equation (\ref{eqn:MasterEqn}) into discrete quantum jumps and evolution via a non-Hermitian Hamiltonian is not unique. In fact, different unravellings may model different measurement processes used to gather information about the internal dynamics of an open quantum system. In particular, the unravellings we are concerned with give us the ability to access phase information that is otherwise obscured by the photocounting method given above. We will consider here an unravelling of the master equation based on the theory of quantum optical quadrature measurements \cite{wiseman_quantum_1993,wiseman_interpretation_1993}.

Instead of considering a single photodetector monitoring the output of the cavity, we consider the output field of the cavity entering a 50/50 beamsplitter, with a strong local oscillator incident on the other input arm of the beamsplitter. The local oscillator is modelled as a classical coherent field with large amplitude $A\in\mathbb{R}$ oscillating with a detuning $\Omega$ from the frequency of the rotating frame and a phase $\theta$. The outputs of the beamsplitter lead to two detectors $D_1$ and $D_2$. The annihilation operators for the fields at the two detectors are 
\begin{align}
\hat{C}_1 &= \sqrt{2\kappa}\hat{a} - Ae^{-i\Omega t}e^{i\theta}, \\
\hat{C}_2 &= i\sqrt{2\kappa}\hat{a} + iAe^{-i\Omega t}e^{i\theta}, 
\end{align}
respectively. In the limit of a very strong local oscillator $A^2\gg \braket{\hat{a}^{\dagger}\hat{a}}$, most of the counts incident on the photodetectors will be due to the local oscillator, with a small number of counts arising due to the internal dynamics of the cavity. Our goal is to use the quantum trajectory approach in the limit $A\rightarrow\infty$ to model a continuous diffusion process that can be cast into the form of a stochastic differential equation for the quantum state. To this end, consider the average number of photocounts into detector $D_1$ over a time $\Delta t$,

\begin{equation}
\bar{N}_1 = \braket{\hat{C}^{\dagger}_1\hat{C}_1}\Delta t = (A^2 - \sqrt{2\kappa}A\braket{e^{i\Omega t}e^{-i\theta}\hat{a} + e^{i\Omega t}e^{-i\theta}\hat{a}^{\dagger} } + 2\kappa\braket{\hat{a}^{\dagger}\hat{a}})\Delta t,
\end{equation}
and the average number of photocounts in detector $D_2$
\begin{equation}
\bar{N}_2 = \braket{\hat{C}^{\dagger}_2\hat{C}_2}\Delta t = (A^2 + \sqrt{2\kappa}A\braket{e^{i\Omega t}e^{-i\theta}\hat{a} + e^{-i\Omega t}e^{i\theta}\hat{a}^{\dagger}} + 2\kappa\braket{\hat{a}^{\dagger}\hat{a}})\Delta t,
\end{equation}
where $\hat{X}_{\theta}(t)=e^{i\Omega t}e^{-i\theta}\hat{a}+e^{-i\Omega t}e^{i\theta}\hat{a}^{\dagger}$ is a quadrature operator. Approaching the limit $A\rightarrow\infty$, where quantum jumps occur at an infinite rate, the quantum dynamics appear as a simultaneous combination of quantum jumps governed by the collapse operators $\hat{C}_{1}$ and $\hat{C}_{2}$, and continous non-unitary evolution governed by the non-Hermitian Hamiltonian (\ref{eqn:nonHermitianHamiltonian}), so the mapping of quantum states from time $t_j$ to $t_{j+1}$ is

\begin{equation}
\ket{\bar{\psi}_{j+1}} = \hat{C}_1\hat{C}_2e^{-i\hat{H}_{\mathrm{eff}}\Delta t_{j+1}}\ket{\bar{\psi}_{j}},
\end{equation}
where the bar over a quantum state denotes the unnormalized form, and $\Delta t_{j+1} = t_{j+1} - t_{j}$. Instead of considering the times where quantum jumps occur at $t_{j}$, $t_{j+1}$ etc., to derive a diffusive form of the quantum dynamics, we consider a coarse grained time interval $\Delta t$ over many quantum jumps, so we have
\begin{align}
\ket{\bar{\psi}(t + \Delta t)} &= \hat{C}_{1}^{N_1}\hat{C}_{2}^{N_2}e^{-i\hat{H}_{\mathrm{eff}}\Delta t}\ket{\bar{\psi}(t)}, \\
&= \left( \sqrt{2\kappa}\hat{a} - Ae^{-i\Omega t}e^{i\theta} \right)^{N_1}\left( i\sqrt{2\kappa}\hat{a} + iAe^{-i\Omega t}e^{i\theta} \right)^{N_2}e^{-i\hat{H}_{\mathrm{eff}}\Delta t}\ket{\bar{\psi}(t)}, \\
&= i^{N_{2}}\left( 1 - \frac{\sqrt{2\kappa}}{A}e^{i\Omega t}e^{-i\theta}\hat{a} \right)^{N_1}\left( 1 + \frac{\sqrt{2\kappa}}{A}e^{i\Omega t}e^{-i\theta}\hat{a}  \right)^{N_2}e^{-i\hat{H}_{\mathrm{eff}}\Delta t}\ket{\bar{\psi}(t)}.
\end{align}  
Applying the binomial expansion to the jump terms and a Taylor expansion of the exponential up to $\mathcal{O}(\Delta t^2)$, the mapping becomes
\begin{equation}\label{eqn:mapping}
\ket{\bar{\psi}(t+\Delta t)} = i^{N_2}\left( 1 + \frac{\sqrt{2\kappa}}{A}(N_1 - N_2)e^{i\Omega t}e^{-i\theta}\hat{a} \right)\left( 1 - i\hat{H}_{\mathrm{eff}}\Delta t \right)\ket{\bar{\psi}(t)}.
\end{equation}
The number of counts incident on the photodetectors $N_{1}$ and $N_{2}$ are interpreted as Poissonian random variables with variances $\Delta N_1 = \Delta N_2 = A\sqrt{\Delta t}$. Over many quantum jumps the central limit theorem may be applied and $N_1$ and $N_2$ may be considered Gaussian random variables. The difference of the counts is then also a Gaussian random variable
\begin{equation}\label{eqn:difference}
N_1 - N_2 = 2\sqrt{2\kappa}A\braket{e^{i\Omega t}e^{-i\theta}\hat{a} + e^{-i\Omega t}e^{i\theta}\hat{a}^{\dagger}}\Delta t + A\Delta W,
\end{equation}
where $\Delta W\in\mathbb{R}$ is a Wiener increment. Substituting (\ref{eqn:difference}) into (\ref{eqn:mapping}) we get
\begin{equation}\label{eqn:finalMapping}
\ket{\bar{\psi}(t + \Delta t)} = \left[ 1 - i\hat{H}_{\mathrm{eff}}\Delta t + \left( 4\kappa\braket{e^{i\Omega t}e^{i\theta}\hat{a} + e^{-i\Omega t}e^{i\theta}\hat{a}^{\dagger}}\Delta t + \sqrt{2\kappa}\Delta W \right)e^{i\Omega t}e^{-i\theta}\hat{a} \right]\ket{\bar{\psi}(t)},
\end{equation}
where the factor $i^{N_2}$ can be discarded as it only contributes an overall phase. In the limit $\Delta t \rightarrow dt$, $\Delta W \rightarrow dW$, $(\Delta W)^2\rightarrow dt$, $(dt)^2\rightarrow 0$, with $\ket{\bar{\psi}(t + \Delta t)} = \ket{\bar{\psi}(t)} + d\ket{\bar{\psi}}$, this may be cast into the form of a stochastic Schr\"{o}dinger equation
\begin{equation}
d\ket{\bar{\psi}} = \left[ -i\hat{H}_{\mathrm{eff}}dt + \left( 4\kappa\braket{e^{i\Omega t}e^{-i\theta}\hat{a} + e^{-i\Omega t}e^{i\theta}\hat{a}^{\dagger}}dt + \sqrt{2\kappa}dW \right)e^{i\Omega t}e^{-i\theta}\hat{a} \right]\ket{\bar{\psi}}.
\end{equation}
When the frequency of the local oscillator is identical to the frequency of the rotating frame (for instance if the probe laser is also used to drive the cavity), the detuning $\Omega$ is zero, leading to the \emph{stochastic Schr\"{o}dinger equation for homodyne detection}
\begin{equation}
d\ket{\bar{\psi}} = \left[ -i\hat{H}_{\mathrm{eff}}dt + \left( 4\kappa\braket{e^{-i\theta}\hat{a} + e^{i\theta}\hat{a}^{\dagger}}dt + \sqrt{2\kappa}dW \right)e^{-i\theta}\hat{a} \right]\ket{\bar{\psi}}.
\end{equation}
On the other hand, if the probing laser is far detuned from the internal dynamics, i.e., $\Omega\gg 2\kappa$, (\ref{eqn:finalMapping}) can be time-averaged over $\Delta t$ to form the \emph{stochastic Schr\"{o}dinger equation for heterodyne detection} \cite{wiseman_interpretation_1993}
\begin{equation}
d\ket{\bar{\psi}} = \left[ -i\hat{H}_{\mathrm{eff}}dt + \left( 2\kappa\braket{\hat{a}^{\dagger}}dt + \sqrt{2\kappa}d\xi \right)\hat{a} \right]\ket{\bar{\psi}},
\end{equation}
where $d\xi$ is a complex Wiener increment.

\bibliography{SMGKP_Switching}